\documentstyle[graphicx,12pt]{article}
\begin{document}

\small
\hoffset=-1truecm
\voffset=-2truecm
\title{\bf The Casimir force between parallel plates in Randall-Sundrum I model}
\author{CHENG Hong-Bo\footnote {E-mail address:
hbcheng@sh163.net}\\
Department of Physics, East China University of Science and
Technology,\\ Shanghai 200237, China\\
The Shanghai Key Laboratory of Astrophysics,\\ Shanghai 200234,
China}

\date{}
\maketitle

\begin{abstract}
The Casimir effect for parallel plates within the frame of
five-dimensional Randall-Sundrum model with two branes is
reexamined. We argue that the nature of Casimir force is repulsive
if the distance between the plates is not extremely tiny, which is
not consistent with the experimental phenomena, meaning that the
Randall-Sundrum I model can not be acceptable. We also point out
that the estimation of the separation between the two branes, by
means of Casimir effect for two-parallel-plate system, is not
feasible, in contrast to another recent study.
\end{abstract}
\vspace{8cm} \hspace{1cm} PACS number(s): 11.10.Kk, 03.70.+k

\newpage

The high-dimensional spacetime theory suggesting that our
observalbe four-dimensional world is a subspace of a higher
dimensional spacetime has a long tradition that was put forward by
Kaluza and Klein more than 80 years ago [1, 2]. The
high-dimensional spacetime models including the dimensionality and
the geometric characteristics of extra dimensions are necessary.
The main motivation for such approaches are to unify the
fundamental interactions in nature. The approaches with additional
dimensions were invoked for providing a breakthrough of
cosmological constant and the hierarchy problems [3-8]. The issues
of high-dimensional spacetimes have their own compactification and
properties of extra dimensions. More theories need developing and
to be realized within the frame with extra dimensions. In the
Kaluza-Klein theory, one extra dimension in our universe was
introduced to be compactified to unify gravity and classical
electrodynamics. The quantum gravity such as string theories or
braneworld scenario is developed to reconcile the quantum
mechanics and gravity with the help of introducing seven extra
spatial dimensions. The new approaches propose that the geometry
of the extra spatial dimensions is responsible for the hierarchy
problem. At first, the large extra dimensions (LED) were
introduced [6-8]. In this model the additional dimensions are flat
and of equal size and the radius of a toroid is limited while the
size of extra space can not be too small, or the hierarchy problem
remains. Another model with warped extra dimensions was also put
forward [9, 10]. A four-dimensional theory compactified on a
$S^{1}/Z_{2}$ manifold with bulk boundary cosmological constants
leading to a stable four-dimensional low energy effective theory,
named Randall-Sundrum (RS) models, suggested that the compact
extra dimension with large curvatures to explain the reason why
the large gap between the Planck and the electroweak scales
exists. In RSI, one of the RS models, there are two 3-branes with
equal opposite tensions and they are localized at $y=0$ and $y=L$
respectively, with $Z_{2}$ symmetry $y\longleftrightarrow-y$,
$L+y\longleftrightarrow L-y$. The Randall-Sundrum model becomes
RSII when one brane is located at infinity, like
$L\longrightarrow\infty$. The standard modelfields and gauge
fields live on the negative tension brane which is visible, while
the positive tension brane with a fundamental scale $M_{RS}$ is
hidden. Employing the additional compactified dimensions and
additional warped dimension can help us to unify the interactions
and resolve the hierarchy problem respectively.

The Casimir effect depends on the dimensionality and topology of
the spacetime [11-20] and has received a great deal of attention
within spacetime models including additional spatial dimensions.
There exists strong influence from the possibility of the
existence, the size and the geometry of extra dimensions on the
Casimir effect, the evaluation of the vacuum zero-point energy.
The precision of the measurements of the attraction force between
parallel plates as well as other geometries has been greatly
improved practically [21-24], leading the Casimir effect to be a
remarkable observable and trustworthy consequence of the existence
of quantum fluctuations. The experimental results clearly show
that the attractive Casimir force between the parallel plates
vanishes when the plates move apart from each other to the very
distant place. In particular it must be pointed out that no
repulsive force appears in this case. Therefore the Casimir effect
for parallel plates can become a window to probe the
high-dimensional Universe and can be used to research on a large
class of related topics on the various models of spaccetimes with
more than four dimensions. More efforts have been made on the
studies. Within the frames of several kinds of spacetimes with
high dimensionality the Casimir effect for various systems has
been discussed. The electromagnetic Casimir effect for parallel
plates in a high-dimensional spacetime has been studied and the
subtraction of the divergences in the Casimir energy at the
boundaries is realized [25, 26]. Some topics were studied in the
high-dimensional spacetime described by Kaluza-Klein theory. It is
shown analytically that the extra-dimension corrections to the
Casimir effect for a rectangular cavity in the presence of a
compactified universal extra dimension are very manifest [27]. It
was also proved rigorously that there will appear repulsive
Casimir force between the parallel plates when the plates distance
is sufficiently large in the spacetime with compactified
additional dimensions, and the higher the dimensionality is, the
greater the repulsive force is, unless the Casimir energy outside
the system consisting of two parallel plates is considered
[27-37]. The research on the Casimir energy within the frame of
Kaluza-Klein theory to explain the dark energy has been performed
and is also fundamental, and a lot of progress have been made
[38-40]. In the context of string theory the Casimir effect was
also investigated [41-44]. Also in the Randall-Sundrum model, the
Casimir effect has been investigated to stabilize the distance
between branes (radon) [45-49]. In particular the evaluation of
the Casimir force between two parallel plates under Dirichlet
conditions has been performed in the Randall-Sundrum models with
one extra dimension [50-52]. The upper limit on the separation
between two branes $kR\leq20$ if the curvature parameter $k$ of
five-dimensional anti de Sitter spacetime ($AdS_{5}$) is equal to
the Planck scale has been obtained in RSI model. It was shown that
the required value for solving the hierarchy problem is $kR\sim12$
and does not conflict with the results like $kR\leq20$ from the
Casimir effect. In the case of RSII model, the influence from
extra dimension on the Casimir force between parallel plates is so
small that it can be neglected. The Casimir effect is an efficient
tool to explore the high-dimensional spacetime.

It is fundamental and significant to re-examine the Casimir effect
for two parallel plates in the Randall-Sundrum I models. Here we
argue that the Casimir force between the two parallel plates could
be repulsive in the five-dimensional RSI background and the
repulsive force results disfavoured by the experimental evidence
will contradict the conclusions from Ref. [50]. Certainly the RSI
model can not be accepted. The main purpose of this paper is to
scrutinize the two-parallel-plate Casimir force in the RSI model
again in order to confirm the nature of the Casimir force,
comparing with the clear and definite experimental results. We
obtain the Casimir force within the RSI model by means of the
zeta-function regularization and discuss the dependence of the
force on the plates gap and the separation of the two branes. Our
conclusions are emphasized in the end.

In this study, we investigate a massless scalar field living in
the bulk in the five-dimensional RSI model of spacetime. Within
the frame the spacetime metric is chosen as,

\begin{equation}
ds^{2}=e^{-2k|y|}g_{\mu\nu}dx^{\mu}dx^{\nu}-dy^{2}
\end{equation}

\noindent where the variable $k$ assumed to be of the order of the
Planck scale governs the degree of curvature of the $AdS_{5}$ with
constant negative curvature. That the extra dimension is
compactified on an orbifold gives rise to the generation of the
absolute value of $y$ in the metric. We follow the procedure of
Ref. [50, 52]. In the five-dimensional spacetime with the
background metric denoted in Eq.(1), the equation of motion for a
massless bulk scalar field $\Phi$ is,

\begin{equation}
g^{\mu\nu}\partial_{\mu}\partial_{\nu}\Phi+e^{2ky}\partial_{y}(e^{-4ky}\partial_{y})=0
\end{equation}

\noindent where $g^{\mu\nu}$ is the usual four-dimensional flat
metric with signature $-2$. The field confining between the two
parallel plates satisfies the Dirichlet boundary conditions
$\Phi(x^{\mu}, y)|_{\partial\Omega}=0$, $\partial\Omega$ positions
of the plates in coordinates \textbf{x}. We can choose the
$y-$dependent part of the field $\Phi(x^{\mu}, y)$ as
$\chi^{(N)}(y)$ in virtue of separation of variables. Having
solved the equation of motion of $\chi^{(N)}(y)$ we obtain their
general expression for the nonzero modes in terms of Bessel
functions of the first and second kind as,

\begin{equation}
\chi^{(N\neq0)}(y)=e^{2ky}(a_{1}J_{2}(\frac{m_{N}e^{ky}}{k})
+a_{2}Y_{2}(\frac{m_{N}e^{ky}}{k}))
\end{equation}

\noindent where $a_{1}$ and $a_{2}$ are arbitrary constants. The
effective mass term for the scalar field denoted as $m_{N}$ can
also be obtained by means of integration out the fifth dimension
$y$. In the case of RSI model, the hidden and visible 3-branes are
located at $y=0$ and $y=\pi R$ respectively, which leads the
Neumann boundary conditions $\frac{\partial\chi^{(N)}}{\partial
y}|_{y=0}=\frac{\partial\chi^{(N)}}{\partial y}|_{y=\pi R}=0$, so
a general reduced equation reads,

\begin{equation}
m_{N}\approx e^{-\pi kR}(N+\frac{1}{4})k\pi=\kappa(N+\frac{1}{4})
\end{equation}

\noindent where

\begin{equation}
\kappa=\pi ke^{-\pi kR}
\end{equation}

\noindent by means of the asymptotic form of the Bessel function
of the first kind $J_{\nu}(x)$ and here we assume $N\gg1$ or
equivalently $\pi kR\gg1$ throughout our work. we must point out
that the effective mass term for the scalar field $m_{N}$ with the
integration out the fifth dimension $y$ holds $m_{N}=0$ when $N=0$
because the first zero of the Bessel function $J_{1}(x)$ vanishes.

The models of the vacuum for parallel plates under the Dirichlet
and modified Neumann boundary conditions as mentioned above in RSI
model can be expressed as,

\begin{equation}
\omega_{nN}=\sqrt{\textbf{k}_{\perp}^{2}+(\frac{n\pi}{a})^{2}+m_{N}^{2}}
\end{equation}

\noindent where

\begin{equation}
\textbf{k}_{\perp}^{2}=\textbf{k}_{1}^{2}+\textbf{k}_{2}^{2}
\end{equation}

\noindent here $\textbf{k}_{1}$ and $\textbf{k}_{2}$ are the wave
vectors in directions of the unbound space coordinates parallel to
the plates surface and $a$ is the distance of the plates. Here $n$
and $N$ represent positive integer. Therefore the total energy
density of the fields in the interior of the system involving two
parallel plates in the RSI model reads,

\begin{eqnarray}
\varepsilon=\int d^{2}\textbf{k}_{\perp}\sum_{n,
N=0}^{\infty'}\omega_{nN} \hspace{9.5cm}\nonumber\\
=-\frac{\sqrt{\pi}}{4}\Gamma(-\frac{3}{2})\kappa^{3}
[E_{2}(-\frac{3}{2};\frac{\pi^{2}}{\kappa^{2}a^{2}},1;0,\frac{1}{4})
-(\frac{\pi}{\kappa a})^{3}E_{1}(-\frac{3}{2};1;\frac{\kappa
a}{4\pi})+\frac{\pi^{3}}{a^{3}}\zeta(-3)-\frac{1}{64}]
\end{eqnarray}

\noindent where the prime means that the term with $n=N=0$ is
excluded and the zeta functions of Epstein-Hurwitz type are
defined by,

\begin{eqnarray}
E_{p}(s;a_{1},a_{2},\cdot\cdot\cdot,a_{p};c_{1},c_{2},\cdot\cdot\cdot,
c_{p})\nonumber\\
=\sum_{\{n_{j}\}=0}^{\infty}(\sum_{j=1}^{p}a_{j}(n_{j}+c_{j})^{2})^{-s}\hspace{1cm}
\end{eqnarray}

\noindent and

\begin{equation}
E_{1}(s;a;c)=\sum_{n=1}^{\infty}(an^{2}+c)^{-s}
\end{equation}

\noindent here $\{n_{j}\}$ stands for a short notation of $n_{1}$,
$n_{2}$, $\cdot\cdot\cdot$, $n_{p}$, $n_{j}$ a nonnegative
integer, and $\zeta(s)=\sum_{n=1}^{\infty}\frac{1}{n^{s}}$ is
Riemann zeta function and
$\zeta_{H}(s,q)=\sum_{n=0}^{\infty}(n+q)^{-s}$ is the Hurwitz zeta
function. Following the approach of Ref.[50], we regularize the
Eq.(8) to obtain the Casimir energy density of parallel plates in
the five-dimensional background governed by RSI model as follow,

\begin{eqnarray}
\varepsilon_{C}=\frac{\pi}{384}\kappa^{3}-\frac{\pi^{4}}{360}\frac{\kappa^{3}}{\mu^{3}}
+\frac{\pi}{32}\frac{\kappa^{3}}{\mu}\sum_{n=1}^{\infty}n^{-2}K_{2}
(\frac{\mu}{2\sqrt{\pi}}n)\hspace{1cm}\nonumber\\
-\frac{1}{2}\frac{\kappa^{3}}{\mu}\sum_{n_{1}=1}^{\infty}\sum_{n_{2}=0}^{\infty}
n_{1}^{-2}(n_{2}+\frac{1}{4})^{2}K_{2}(2\mu
n_{1}(n_{2}+\frac{1}{4}))
\end{eqnarray}

\noindent and

\begin{equation}
\mu=\kappa a=\pi kae^{-\pi kR}
\end{equation}

\noindent and here $K_{\nu}(z)$ is the modified Bessel function of
the second kind. The terms with series converge very quickly and
only the first several summands need to be taken into account for
numerical calculation to further discussion. We discuss the
Casimir energy density in the limiting case that the plates
separation $a$ is large enough,

\begin{equation}
\varepsilon_{C}(\mu\gg1)=\frac{\pi}{6}\kappa^{3}\zeta_{H}(-3,\frac{1}{4})
-\frac{\pi}{384}\kappa^{3}<0
\end{equation}

\noindent which means that the sign of the Casimir energy keeps
negative. Certainly we continue focusing on the Casimir force
which is given by the derivative of the Casimir energy with
respect to the plate distance. According to the Casimir energy
density denoted as (11), the Casimir force per unit area of plates
becomes,

\begin{eqnarray}
f_{C}=-\frac{\partial\varepsilon_{C}}{\partial a}\hspace{11cm}\nonumber\\
=-\frac{1}{2}\frac{\kappa^{4}}{\mu^{2}}\sum_{n_{1}=1}^{\infty}
\sum_{n_{2}=0}^{\infty}n_{1}^{-2}(n_{2}+\frac{1}{4})^{2}
K_{2}(2\mu n_{1}(n_{2}+\frac{1}{4}))\hspace{4.5cm}\nonumber\\
-\frac{1}{2}\frac{\kappa^{4}}{\mu}\sum_{n_{1}=1}^{\infty}
\sum_{n_{2}=0}^{\infty}n_{1}^{-1}(n_{2}+\frac{1}{4})^{3}
[K_{1}(2\mu n_{1}(n_{2}+\frac{1}{4}))+K_{3}(2\mu
n_{1}(n_{2}+\frac{1}{4}))]\hspace{0.5cm}\nonumber\\
+\frac{\pi}{32}\frac{\kappa^{4}}{\mu^{2}}\sum_{n=1}^{\infty}n^{-2}
K_{2}(\frac{\mu}{2\sqrt{\pi}}n)
+\frac{\sqrt{\pi}}{128}\frac{\kappa^{4}}{\mu}\sum_{n=1}^{\infty}
n^{-1}[K_{1}(\frac{\mu}{2\sqrt{\pi}}n)+K_{3}(\frac{\mu}{2\sqrt{\pi}}n)]
\hspace{0.5cm}\nonumber\\
-\frac{\pi^{4}}{1080}\frac{\kappa^{4}}{\mu^{4}}\hspace{10.5cm}
\end{eqnarray}

\noindent It is enough to make the first several summands because
of quickly convergent series in the expression. If the
dimensionless variable $\mu\gg1$, which means that the gap between
two plates is much larger than the separation between the two
branes in the RSI model, the Casimir force vanishes,

\begin{equation}
\lim_{\mu\longrightarrow\infty}f_{C}=0
\end{equation}

\noindent which is not in conflict with the experimental evidence.
We have to perform the burden and surprisingly difficult
calculation in order to scrutinize the nature of the
two-parallel-plate Casimir force in the RSI model within a wider
region. The dependence of the Casimir force on the variable
$\mu=\kappa a$ is potted in the Figure 1. We calculate the Casimir
force expression (14) to find that there must exist a special
value of variable $\mu$ denoted as $\mu_{0}=0.156$. When
$\mu<\mu_{0}$, the sign of the Casimir force between the two
parallel plates is negative, which means that the plates attract
each other. When the plates separation is sufficiently large to
keep $\mu>\mu_{0}$, the nature of the Casimir force is repulsive.
Although the Casimir force is equal to the zero as the distance
between the plates approaches to the infinity, it should be
pointed out that the force keeps repulsive during the process as
$\mu>\mu_{0}$. It is clear that our results are different from
those of Ref. [50]. The results that the Casimir force within the
system containing two parallel plates is repulsive are not
consistent with the experimental phenomenon. In this system no
repulsive Casimir force appears according to the measurements. It
should also be emphasized that the value of special parameter
$\mu_{0}$ can be different for different kinds of fields referring
to more complicated boundary conditions than the case of scalar
field we consider here, but the repulsive Casimir force must
generate as the plates are sufficiently far away from each other.
It is necessary to make some estimations for the sake of
comparison. In Ref. [50] the Casimir force was plotted within the
range of plate separation mainly from $0.5\times10^{-6}m$ to
$2\times10^{-6}m$ by means of comparison of the Casimir force in
RSI model with the ones in the case of the standard Casimir force
supported by measurements without RS contribution to obtain the
upper bound $kR\leq20$ while the $AdS_{5}$ curvature scale $k$ is
set to the $10^{16}GeV$ or Planck scale $10^{19}GeV$. In Ref. [50]
the Casimir force was drawn in Figures when the plates distance
belongs to the range mainly from $0.5\times10^{-6}m$ to
$3\times10^{-6}m$. Having substituted the results from Ref. [50]
like upper bound $kR\leq20$ and the values of the $AdS_{5}$
curvature scale $k$ and plate distance like
$0.5\times10^{-6}m\leq\mu\leq3\mu m$ which were employed by Ref.
[50] into the relation between the plate distance and radion
denoted in (12), we give rise to the range of the dimensionless
variable $\mu\in[41.077, 164.308]$. Certainly the values of
dimensionless variable in the case of Ref. [50] are much larger
than the special value $\mu_{0}$ that we discover above and the
Casimir force should be positive which lead the two parallel
plates to move apart. In addition we research on the restriction
on the plate gap when we set $kR\sim12$ the required value for
solving the hierarchy problem and $k$ to the Planck scale.
According to the definition of $\mu$ like (12) and the special
value $\mu_{0}=0.156$, we find that the Casimir force between
plates is attractive only when the restriction on plate separation
is $a<2.2\times10^{-20}m$ or the force becomes repulsive, which is
not consistent with the experimental results. It should be pointed
out that the equation is valid asymptotically for $N\gg1$ although
the reduced equation (4) for the effective mass of the scalar bulk
field is expressed as an approximation. According to the
properties of Bessel functions of the first and second kind, the
error is about $3\%$ when $N=1$ and the error is $0.3\%$ and
$0.1\%$ for $N=2$ and $N=3$ respectively, etc., displaying that
the error drops very quickly with increasing $N$. The deviation
from the approximation can not change our conclusion, so the
repulsive Casimir force denoted as positive magnitude of $f_{C}$
will appear inevitably when the plates distance is sufficiently
large.

In conclusion, there must appear the repulsive force between two
parallel plates in the five-dimensional Randall-Sundrum model with
two branes if the plate separation is not extremely tiny and the
results that the Casimir force is repulsive conflict with the
experimental evidence. We come to a different conclusion from
those of Ref. [50]. Having discussed the Casimir force between
parallel plates in the frame of RSI model in detail, we reveal
that the Casimir force always remains repulsive as the plate
separation  is larger than a very small quantity which is equal to
$10^{-20}m$ approximately while $kR\sim 12$, the required value
for solving the hierarchy problem and $k$ is set to be the Planck
scale, but the experimental evidence confirms that no repulsive
Casimir force appears in this case. Of course it is impossible to
estimate the branes distance of RSI model by means of Casimir
force for standard two-parallel-plate device. Although we are
limited here in the case of massless scalar field obeying the
Dirichlet boundary condition for simplicity and comparison, the
special variable $\mu_{0}$ will be different for different fields
with different kinds of boundary conditions, but the special
parameter $\mu_{0}$ must exist. If the dimensionless variable
$\mu$ showing the relation between the separations of plates and
branes is larger than the special value $\mu_{0}$ which is very
small, the Casimir force must be repulsive which is excluded by
the experiment, which mean that the Randall-Sundrum I model can
not be acceptable. The topics about the Casimir force in the
five-dimensional Randall-Sundrum model needs to be developed
further and related topics also need further research.

\vspace{3cm}

\noindent\textbf{Acknowledgement}

This work is supported by NSFC No. 10875043 and is partly
supported by the Shanghai Research Foundation No. 07dz22020.

\newpage
\begin{figure}
\setlength{\belowcaptionskip}{10pt} \centering
  \includegraphics[width=15cm]{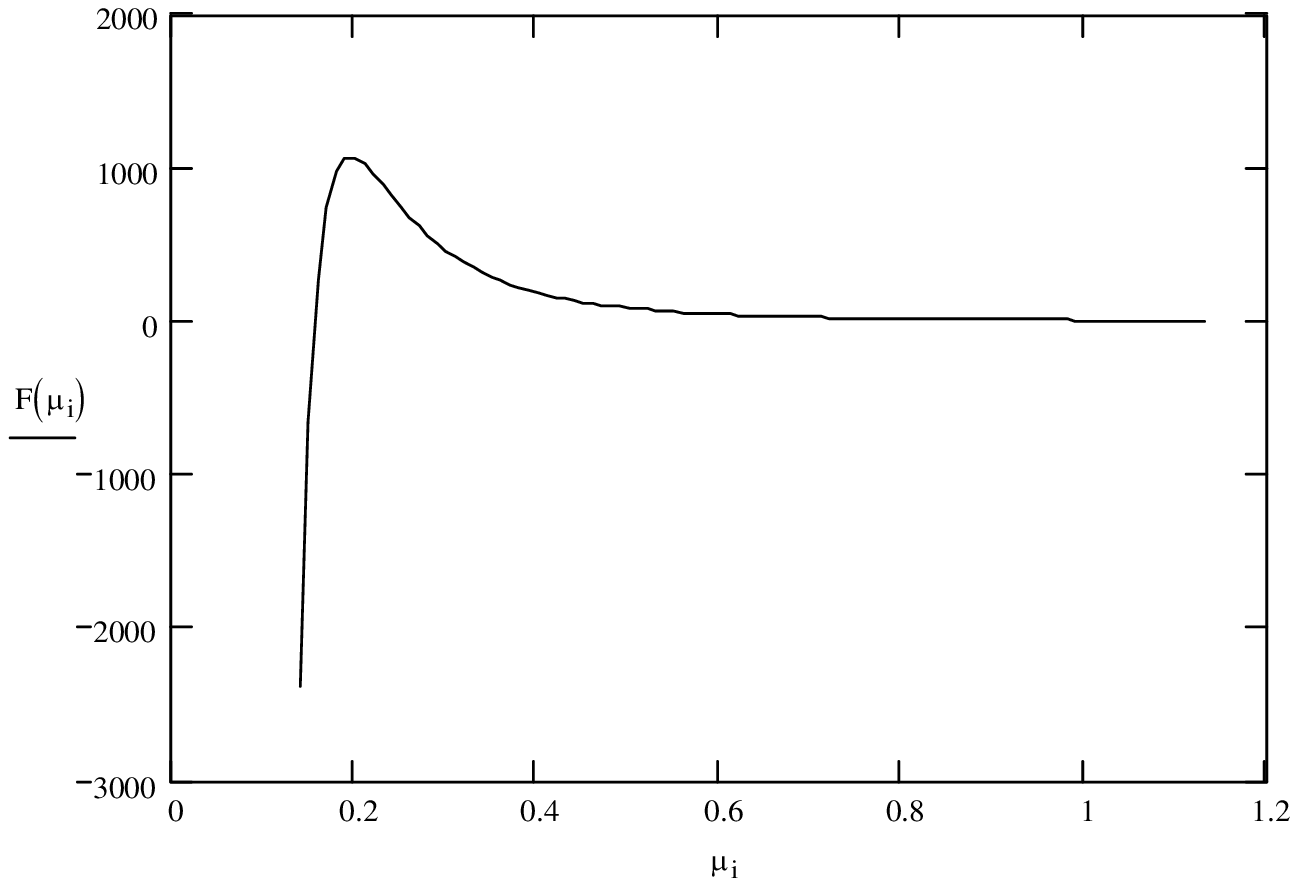}
  \caption{The Casimir force per unit area in unit of $\kappa^{4}$
  between parallel plates versus the dimensionless variable denoted as $\mu=\kappa a$}
\end{figure}

\end{document}